\documentclass[11pt,a4paper]{article}

\newcommand{\Z}{\mathbb{Z}}
\newcommand{\C}{\mathbb{C}}

\newcommand{\D}{\mathcal{D}}
\newcommand{\M}{\mathcal{M}}
\newcommand{\N}{\mathcal{N}}

\newcommand{\Hom}{\mathrm{Hom}}
\newcommand{\Tr}{\mathrm{Tr}}
\newcommand{\tr}{\mathrm{tr}}

\newcommand{\U}{\mathrm{U}}
\newcommand{\SU}{\mathrm{SU}}
\newcommand{\SO}{\mathrm{SO}}
\newcommand{\GL}{\mathrm{GL}}

\newcommand{\beq}{\begin{eqnarray}}
\newcommand{\eeq}{\end{eqnarray}}
\newcommand{\ba}{\left( \begin{array}}
\newcommand{\ea}{\end{array} \right)}
\newcommand{\be}{\begin{equation}}
\newcommand{\ee}{\end{equation}}
\newcommand{\bea}{\begin{eqnarray}}
\newcommand{\eea}{\end{eqnarray}}
\newcommand{\beann}{\begin{eqnarray*}}
\newcommand{\eeann}{\end{eqnarray*}}

\newcommand{\hs}[1]{\hspace{#1 mm}}
\newcommand{\diag}{\mathrm{diag}}
\def\p{\partial}

\usepackage{bm,bbm,amsfonts,mathrsfs}
\usepackage{graphicx}
\usepackage{hyperref}
\makeatletter

\@addtoreset{equation}{section}
\makeatother

\date{\today}

\begin{document}

\begin{titlepage}

\renewcommand{\thefootnote}{\fnsymbol{footnote}}

\begin{flushright}
IFUP-TH/2012-06 \\
RIKEN-MP-46 

\end{flushright}

\vskip5em

\begin{center}
 {\LARGE {\bf 
 Vortex counting from field theory
 }}

 \vskip3em

 {\sc Toshiaki Fujimori},\footnote{E-mail address:
 \href{mailto:toshiaki.fujimori@pi.infn.it}
 {\tt toshiaki.fujimori@pi.infn.it}}$^{1,2}$
 {\sc Taro Kimura},\footnote{E-mail address: 
 \href{mailto:tkimura@ribf.riken.jp}
 {\tt tkimura@ribf.riken.jp}}$^{3,4}$
 {\sc Muneto Nitta}\footnote{E-mail address:
 \href{mailto:nitta@phys-h.keio.ac.jp}
 {\tt nitta@phys-h.keio.ac.jp}}$^5$
 and
 {\sc Keisuke Ohashi}\footnote{E-mail address: 
 \href{mailto:ohashi@gauge.scphys.kyoto-u.ac.jp}
 {\tt ohashi@gauge.scphys.kyoto-u.ac.jp}}$^6$

 \vskip2em

$^1${\it INFN, Sezione di Pisa, Largo B. Pontecorvo, 3, Ed. C, 56127
 Pisa, Italy}\\ \vskip.2em
$^2${\it Department of Physics, ``E. Fermi'', University of Pisa, \\
 Largo B.~Pontecorvo, 3, Ed.~C, 56127 Pisa, Italy}\\ \vskip.2em
$^3${\it Department of Basic Science, University of Tokyo, 
 Tokyo 153-8902, Japan}\\ \vskip.2em
$^4${\it Mathematical Physics Laboratory, RIKEN Nishina Center, Saitama
 351-0198, Japan}\\ \vskip.2em
$^5${\it Department of Physics, and Research and Education Center for
 Natural Sciences,\\ Keio University, Hiyoshi 4-4-1, Kanagawa 223-8521,
 Japan}\\ \vskip.2em
$^6${\it Department of Physics, Kyoto University, Kyoto 606-8502, Japan}

 \vskip3em

% \today
\end{center}

 \vskip2em

\begin{abstract}
The vortex partition function in 2d $\mathcal N=(2,2)$ $\U(N)$ 
gauge theory is derived from the field theoretical
point of view by using the moduli matrix approach.
The character for the tangent space at each moduli space fixed point is
written in terms of the moduli matrix, and then the vortex partition
function is obtained by applying the localization formula.
We find that dealing with the fermionic zero modes is crucial 
to obtain the vortex partition function 
with the anti-fundamental and adjoint matters 
in addition to the fundamental chiral multiplets.
The orbifold vortex partition function is 
also investigated from the field theoretical point of view.
\end{abstract}

\end{titlepage}

\tableofcontents

\setcounter{footnote}{0}

\section{Introduction}\label{sec:intro}

The gauge theory partition function plays an essential role in
non-perturbative aspects of supersymmetric gauge theory
\cite{Nekrasov:2002qd,Nekrasov:2003rj}.
It is directly given by performing path integral for a certain
supersymmetric theory \cite{Moore:1997dj}, and correctly provides its low energy
dynamics \cite{Seiberg:1994rs,Seiberg:1994aj}.
Furthermore it is shown that the instanton partition function is directly
interpreted as the conformal block of the two dimensional Liouville field
theory \cite{Alday:2009aq}.
It can be also regarded as a consequence of the M-brane
compactifications \cite{Witten:1997sc,Gaiotto:2009we}.

Recently partition functions have been provided for the low dimensional
gauge theories by performing vortex counting
\cite{Shadchin:2006yz,Dimofte:2010tz,Yoshida:2011au,Bonelli:2011fq,Bonelli:2011wx}, 
where non-Abelian vortices in $\U(N)$ gauge theories 
\cite{Hanany:2003hp,Auzzi:2003fs,Tong:2005un,Eto:2006pg,Shifman:2007ce,Tong:2008qd} 
play roles of instantons in two dimensions.
In addition the moduli space volume itself is also investigated by using
the localization formula \cite{Miyake:2011yr}.
They are mainly based on Hanany-Tong's approach \cite{Hanany:2003hp}
in which a D-brane construction is used 
to describe the vortex moduli space, 
or on BPS equations themselves.
On the other hand, it is shown that, although Hanany-Tong's approach can
capture the global structure of the vortex moduli space, the local
structure is not correctly treated: the metric of the moduli
space is different from the result obtained from the purely
field theoretical method \cite{Hanany:2003hp,Fujimori:2010fk}.
Thus it is important to check its consistency by investigating
non-perturbative aspects, e.g. a partition function, from the field
theoretical point of view.

Such a non-trivial consequence of the gauge theory is also discussed for
the four dimensional orbifold theory, for example, the instanton counting
\cite{Fucito:2004ry}, the AGT relation
\cite{Belavin:2011pp,Nishioka:2011jk,Bonelli:2011jx,Belavin:2011tb,Bonelli:2011kv,Wyllard:2011mn},
the matrix model description \cite{Kimura:2011zf,Kimura:2011gq} and so
on.
Recently the low dimensional orbifolds with respect to the vortex moduli
space is investigated \cite{Kimura:2011wh}.
The orbifold vortex partition function is given in \cite{Zhao:2011ke},
but it is again based on Hanany-Tong's approach.
Thus we should reconsider it in a field theoretical manner as well as
the vortex partition functions in the standard two dimensional space.

In this paper we apply the moduli matrix approach 
to study the vortex moduli space 
\cite{Eto:2005yh,Eto:2006pg,Eto:2006uw,Eto:2006cx,Eto:2006db,Fujimori:2010fk,Eto:2011pj}.
This is just a purely field theoretical perspective, 
which can systematically deal with solutions of the BPS equations.
The fixed points in the moduli space with respect to the isometry, 
which is coming from the symmetry of the gauge theory, 
are completely classified in terms of the moduli matrix.
Thus we can write down the character for the tangent space
through the moduli matrix method, 
and then obtain the vortex partition function.

This paper is organized as follows.
In section \ref{sec:Kahler}, we review the standard derivation of the
vortex partition function through the K\"ahler quotient, 
on which Hanany-Tong's approach is based.
This is parallel to the ADHM construction of instantons, 
and thus we can obtain the partition function 
in a quite similar manner to the instanton theory.
In section \ref{sec:moduli_mat}, 
we investigate the vortex partition function 
from the field theoretical point of view.
We discuss the structure of the vortex moduli space 
for two dimensional $\mathcal N=(2,2)$ $\U(N)$ gauge theories, 
especially the fixed points of the torus action and 
the corresponding tangent spaces in the moduli space.
We write down the character for each tangent space 
in terms of the moduli matrices, 
and derive the vortex partition functions for the cases with 
the anti-fundamental and adjoint matters 
in addition to $N_{\rm F}=N$ fundamental chiral multiplets.
We find that dealing with the fermionic zero modes is crucial 
to obtain the vortex partition function 
with the anti-fundamental and adjoint matters 
in addition to the fundamental chiral multiplets.
In section \ref{sec:orbifold}, 
we extend our result to the orbifold theory on $\C/\Z_n$.
We derive the consistency conditions for the moduli matrices, 
and then obtain the orbifold partition function 
by considering the tangent space character at each fixed point.
In section \ref{sec:summary}, 
we conclude this paper with some remarks and discussions.

\section{The K\"ahler quotient method}
\label{sec:Kahler}

The vortex partition functions have been obtained 
in a similar way to the case of instantons \cite{Nekrasov:2002qd} 
by utilizing the K\"ahler quotient constructions.
In this section, let us review the K\"ahler quotient method to fix the
notation we use in the following discussions.

First, we consider 2d $\mathcal N = (2,2)$ 
(4d $\mathcal N =1$)
$\U(N)$ gauge theory with $N$ fundamental chiral multiplets, 
whose bosonic Lagrangian is given by 
\begin{equation}
 \mathcal L_b \, = \, \tr \bigg[ - \D_\mu H (\D^\mu H)^\dagger - \frac{1}{2g^2} F_{\mu \nu} F^{\mu \nu} - \frac{g^2}{4} \left( H H^\dagger - v^2 {\mathbf 1}_N \right)^2 \bigg]
  \label{boson_action01}
\end{equation}
where the adjoint scalar fields in the 2d vector multiplets 
are omitted because they do not contribute to 
the BPS equations obtained below.
Here $H$ is an $N$-by-$N$ matrix, on which 
the color and flavor symmetry act in the following way:
\beq
H \rightarrow U_C H U_F, \hs{10} U_C \in \U(N)_C, \hs{5} U_F \in \U(N)_F.
\eeq
The BPS vortex equations for this Lagrangian are given by 
\cite{Hanany:2003hp,Auzzi:2003fs}
\begin{equation}
 \left(
  \mathcal{D}_1 + i \mathcal{D}_2
 \right) H = 0, \qquad
 F_{12} + \frac{g^2}{2}
 (v^2 {\mathbf 1}_N - HH^\dag) = 0 .
 \label{bosonic_BPS_eq}
\end{equation}
The vortex moduli space is given by the space of the BPS solutions 
with a fixed vortex number $k$ defined by
\beq
k \equiv - \frac{1}{2\pi} \int_{\C} F.
\eeq

A $k$-vortex solution for this $\U(N)$ gauge theory is associated with 
a pair of matrices $(B,I)$ satisfying \cite{Hanany:2003hp},
\begin{equation}
 [ B, B^\dag] + I I^\dag = r {\mathbf 1}_{k},
\end{equation}
where $r$ stands for the Fayet-Illiouplos (FI) parameter on the vortex
worldvolume.\footnote{This FI parameter is related to 
the coupling constant of the original $\U(N)$ gauge theory 
as $r = 4\pi/g^2$.}
We have $B \in \Hom(V,V)$, $I \in \Hom(W,V)$ for two vector spaces $V$
and $W$, whose dimensions are interpreted as the winding number and the
rank of the gauge group, dim~$V=k$ and dim~$W=N$, respectively.
Thus the moduli space is given by
\begin{equation}
 \M_{N,k} \cong
  \left\{
   (B,I) \Big| [ B, B^\dag] + I I^\dag = r {\mathbf 1}_{k}
  \right\} / \U(k).
  \label{moduli_sp1}
\end{equation}
The $\U(k)$ symmetry acts on these data as follows,
\begin{equation}
 (B,I)
  \longrightarrow 
  (g B g^{-1}, g I), \qquad
  g \in \U(k).
\end{equation}
Note that there is another representation of the moduli space,
\begin{equation}
 \M_{N,k} \cong \left\{ (B,I) \right\} / \hs{-1} / \GL(k,\C),
\end{equation}
where the quotient denoted by the double slash $/ \hs{-1} /$ means 
that points at which the $GL(k,\C)$ action is not free 
should be removed so that the group action is free at any point.

Let us consider the isometry $\U(1) \times \U(1)^{N-1}$, 
which acts on the quotient as
\begin{equation}
 (B,I) \longrightarrow (T_{\epsilon} B, I T_{a}^{-1}),
  \label{isometry_KQ}
\end{equation}
where we have denoted the torus action 
as $T_\epsilon = e^{i\epsilon}$ and 
$T_a= \mathrm{diag} (e^{ia_1}, \cdots, e^{ia_N})$.
They are coming from
the global symmetry of the system $\SO(2) \times \SU(N)$: 
the former is the spatial rotation and 
the latter is the color-flavor diagonal symmetry.
Here $a_l$ satisfy $\sum_{l=1}^N a_l = 0$ 
and are called the twisted mass parameters.
The color-flavor symmetry is broken to $\U(1)^{N-1}$ 
due to this twisted mass parameters.

The fixed points in the vortex moduli space are labeled by 
an $N$-tuple of one dimensional partitions, 
which just consists of $N$ integral entries,
\begin{equation}
 \vec{k} = (k_1, \cdots, k_N).
  \label{fixed_pt}
\end{equation}
The vortex number is given by $k = k_1 + \cdots + k_N$.\footnote{
The same decomposition has been made for 
$\SU(N)$-orbits of the vortex moduli space \cite{Eto:2010aj}.
}
We then obtain the character of the vector space 
at the fixed point specified  by $\vec{k}$ as
\begin{equation}
 \chi(V) = \sum_{l=1}^N \sum_{i=1}^{k_l} T_{a_l} T_\epsilon^{i-1}, \qquad
 \chi(W) = \sum_{l=1}^N T_{a_l}.
\end{equation}
Therefore, the character for the tangent space at the fixed point is given by
\begin{eqnarray}
 \chi (T_{\vec{k}} \M) & = & - ( 1 - T_\epsilon ) \, \chi(V^*) \times \chi(V) + \chi(W^*)  \times \chi(V)
  \nonumber \\
 & = & \sum_{l,m}^N \sum_{i=1}^{k_m} T_{a_{ml}} T_\epsilon^{-k_l+i-1} 
~=~ \sum_{l,m}^N \sum_{i=1}^{k_m} e^{i a_{ml} + i(- k_l + i - 1)\epsilon},
  \label{character}
\end{eqnarray}
where we have denoted $a_{ml} = a_m - a_l$.
Then, the vortex partition function for $\U(N)$ gauge theory 
with $N$ fundamental chiral multiplets is given 
by applying the localization formula \cite{Bruzzo:2002xf}: 
it can be found by replacing the sum by the products over the weights as
\begin{equation}
 Z_{\vec{k}} = \prod_{l,m}^N \prod_{i=1}^{k_m} 
  \frac{1}{a_{ml} + (- k_l + i - 1) \epsilon}.
  \label{part_func}
\end{equation}
The number of products in this partition function is $Nk$, which is just
the dimension of the moduli space, dim$_\C \M_{N,k}=Nk$.

It is expected that we can also deal with the cases with the
anti-fundamental and adjoint matters in a similar manner.
In order to apply the method discussed above, we have to
first derive the corresponding K\"ahler quotient to such theories.%
\footnote{The models with the anti-fundamental and adjoint matters
have been so far investigated by using the reduction from 
the ADHM data for instantons \cite{Bonelli:2011fq,Bonelli:2011wx}.}
In the following, we show another derivation of the vortex partition
function without using the K\"ahler quotient method.
We can directly obtain such a partition function from the field theory.

\section{The moduli matrix method}
\label{sec:moduli_mat}

In this section, we show how the partition function (\ref{part_func}) is
derived from the field theoretical viewpoint without using the K\"ahler
quotient.
To obtain the partition by applying the localization formula, 
all we have to do is to derive the character 
of the torus action at each fixed point in the vortex moduli space.
We now explicitly derive this character in terms of the moduli matrix.

\subsection{$\U(N)$ gauge theory with $N$ fundamental matters}
\label{sec:bosonic}

We first consider the bosonic zero modes 
for the case with $N$ fundamental chiral multiplets.
The solution of the BPS equation (\ref{bosonic_BPS_eq}) 
can be written in terms of a holomorphic matrix $H_0(z)$, 
which we call the moduli matrix.
The first equation in (\ref{bosonic_BPS_eq}) 
can be solved as \cite{Isozumi:2004vg,Eto:2005yh,Eto:2006pg} 
\begin{equation}
 H(z,\bar{z}) = v \, S^{-1}(z,\bar{z}) H_0(z), \qquad
 A_{\bar{z}} = \frac{1}{2} \left( A_1 + i A_2 \right) = -i S^{-1}(z,\bar{z}) \partial_{\bar{z}} S(z,\bar{z}), 
 \label{bc03}
\end{equation}
where $z=x_1+ix_2$ is the complex coordinate of the two dimensional space $\C$.
The rank of $H_0(z)$ is $N$ for $\U(N)$ gauge theory.
Then the second BPS equation can be recast into 
a equation for $S \in \GL(N,\C)$, 
which has a unique solution (up to gauge transformations) 
for a given $H_0(z)$ \cite{Lin:2011zzf}.
Therefore, all moduli parameters are contained in the moduli matrix $H_0(z)$. 
This construction is invariant under the so-called ``$V$-transformation''
\begin{equation}
H_0(z) \rightarrow V(z) H_0(z), \hs{10} S(z,\bar{z}) \rightarrow
V(z) S(z,\bar{z}),
\end{equation}
with $V(z) \in \GL(N, \C)$ being holomorphic with respect to $z$.
Since the original fields $(H,A_{\bar z})$ are 
invariant under this transformation,  
we can define the following equivalence relation of the moduli matrix:
\beq
H_0(z) \sim V(z) H_0(z).
\label{eq:v-equiv}
\eeq
Since the energy (action) of $k$-vortex configurations is given by
\begin{equation}
 T = 2 \pi v^2 k
   = - i \frac{v^2}{2} \oint (dz \partial_z - d \bar{z} \p_{\bar{z}}) 
\log |\det H_0(z)|^2,
\end{equation}
the determinant of the moduli matrix for $k$-vortex solutions 
takes the form
\begin{equation}
 \det H_0(z) = \prod_{i=1}^k (z - z_i) . \label{eq:detH_0}
\end{equation}
Here $k$ is the number of vortices on the complex plane $\C$ and
$z_i$ parametrize vortex positions.
Thus the vortex moduli space is represented in terms of the moduli matrix as
\begin{equation}
\mathcal{M}_{N,k} \cong 
\left\{ H_0(z) \Big| \det H_0(z) = \mathcal{O}(z^k) \right\} / \hs{-1} / \{ \mbox{$V$-transformations} \}.
\end{equation}

Next, let us study the fixed points in the vortex moduli space 
in order to calculate the character of the torus action.
Let $H_0^{\vec{k}}(z)$ be the moduli matrix corresponding to the fixed
point with $\vec{k}=(k_1, \cdots, k_N)$, which can be represented as
\begin{equation}
 H_0^{\vec{k}}(z) = \mathrm{diag}(z^{k_1}, \cdots, z^{k_N}).
\label{eq:fixed}
\end{equation}
By using an appropriate $V$-transformation, 
we can see that $H_0^{\vec{k}}(z)$ is invariant 
under the torus action $T_\epsilon =e^{i\epsilon}$ and 
$T_a = \mathrm{diag} (e^{ia_1}, \cdots, e^{ia_N})$,  
\beq
H_0^{\vec k}(z) ~\rightarrow~ V_{\vec k} H_0^{\vec k}(T_\epsilon z) T_a = H_0^{\vec k}(z),
\label{eq:torus}
\eeq
with the corresponding $V$-transformation,
\begin{equation}
 V_{\vec{k}} = \mathrm{diag} ( e^{-i(k_1\epsilon + a_1)}, \cdots,
 e^{-i(k_N \epsilon + a_N)}).
 \label{torus_and_V}
\end{equation}
In order to study the action of the torus action for the tangent space, 
let us consider the neighborhood around the fixed point parametrized 
by a small deviation $\delta H_0(z)$
\begin{equation}
 H_0(z) \approx H_0^{\vec{k}}(z) + \delta H_0(z),
\end{equation}
which obeys the following infinitesimal version of the equivalence relation
(\ref{eq:v-equiv})
\begin{equation}
 \delta H_0(z) \sim \delta H_0(z) + \delta V(z) H_0^{\vec{k}}(z).
\end{equation}
Since $\delta H_0(z)$ is an arbitrary $N$-by-$N$ matrix whose components
are polynomials of $z$, the vector space of all $\delta H_0(z)$ can be
written as
\beq
\{ \delta H_0 \} ~\cong~ \mathbb C^N \otimes \underbrace{\vphantom{\big(} (\mathbb C[z] \oplus \cdots \oplus \mathbb C[z])}_{N},
\eeq
where $\mathbb C[z]$ is the set of polynomials. 
The vector spaces $\mathbb C^N$ and 
$\mathbb C[z] \oplus \cdots \oplus \mathbb C[z]$
correspond to the rows and columns of $\delta H_0(z)$, respectively. 
On the other hand, the space of all infinitesimal $V$-transformation 
can be written as
\beq
\{ \delta V H_0^{\vec k} \} ~\cong~ \mathbb C^N \otimes (I_{k_1}[z] \oplus \cdots \oplus I_{k_N}[z]),
\eeq
where $I_k[z]$ is the set of polynomials which are multiples of $z^k$. 
Therefore, the tangent space is given by
\beq
T_{\vec k} \mathcal M &\cong& \mathbb C^N \otimes (\mathbb C[z]/I_{k_1}[z] \oplus \cdots \oplus \mathbb C[z]/I_{k_N}[z]) \nonumber \\
&\cong& \mathbb C^N \otimes (P_{k_1}[z] \oplus \cdots \oplus P_{k_N}[z]),
\eeq
where $P_k[z]$ is the set of polynomials whose degrees are less than $k$. 
Since the torus action on $\delta H_0$ is written as
\beq
\delta H_0(z) ~\rightarrow~ V_{\vec k} \delta H_0(T_\epsilon z) T_a ,
\eeq
the characters of the torus action on $\mathbb C^N$ and $P_{k_1}[z]
\oplus \cdots \oplus P_{k_N}[z]$ are given by
\beq
\chi(\mathbb C^N) = \Tr \left[ V_{\vec k}(z) \right] = \sum_{l=1}^N (T_\epsilon^{k_l} T_{a_l})^{-1}, \qquad
\chi(P_{k_1}[z] \oplus \cdots \oplus P_{k_N}[z]) = \sum_{l=1}^N T_{a_l} \sum_{i=1}^{k_l} T_\epsilon^{i-1}.
\nonumber \\
\eeq
Therefore, the character of the torus action on $T_{\vec k} \mathcal M$ is
\beq
\chi(T_{\vec k} \mathcal M) &=& \chi(\mathbb C^N) \times \chi(P_{k_1}[z] \oplus \cdots \oplus P_{k_N}[z]) \nonumber \\
&=& \sum_{l=1}^N \sum_{m=1}^N \sum_{i=1}^{k_m} T_{a_{ml}} T_\epsilon^{-k_l + i - 1}.
\eeq
This is consistent with the result from the K\"ahler quotient shown in
(\ref{character}).

Based on the discussion above, we then show an easy way to extract the
character from the moduli matrix.
Each component of the deviation part of the moduli matrix can be represented as
\begin{eqnarray}
 \left(\delta H_0\right)_{lm} & = & \sum_{j=1}^{k_m} c_{lm,j} z^{j-1} .
\end{eqnarray}
Here the number of parameters is the same as the dimension of the moduli
space, dim$_\C\M_{N,k}=Nk$.
Thus it is natural to interpret them as coordinates of (the tangent space of)
the moduli space, which are symbolically denoted as 
$\phi^i~(i=1, \cdots, Nk)$.
Indeed the K\"ahler potential of the moduli space is written down in
terms of these parameters of the moduli matrix \cite{Eto:2006db}.
The isometry, corresponding to (\ref{eq:torus}), 
acts on the coordinates as
\begin{equation}
 c_{lm,j} \longrightarrow e^{i (a_m - a_l)+i(-k_l+j-1)\epsilon}
 c_{lm,j}.
 \label{prefactor}
\end{equation}
These factors coincide with the contribution to the character of
the tangent space (\ref{character}).
Actually, we can extract the character of the tangent space from them.
Eq.~(\ref{prefactor}) shows the torus action on these
coordinates $\phi^{i=(l,m,j)} = c_{lm,j}$ is not only linear 
but also diagonal, that is, the parameters transform as 
$\phi^i \rightarrow {(\mathcal{T}_{\vec k})^i}_j \phi^j$ 
with a diagonal matrix $\mathcal{T}_{\vec k}$.
Its eigenvalues are $e^{i(-k_l+j-1)+i(a_m-a_l)}$, and thus the
character is simply given by
\begin{equation}
 \chi (T_{\vec{k}} \M)
  = \Tr \left[\mathcal{T}_{\vec{k}}\right]
  = \sum_{l,m}^N \sum_{j=1}^{k_m} e^{i(a_m-a_l)+i(-k_l+j-1)\epsilon} .
\end{equation}
The trace is taken at the fixed point labeled by the partition $\vec{k}$.
This is consistent with the result from the K\"ahler quotient shown in
(\ref{character}).

Although we have discussed only the bosonic zero modes in this subsection, 
there also exist fermionic zero modes in the BPS vortex backgrounds. 
For 1/2 BPS vortices in $\mathcal N = (2,2)$ theories, 
there are two conserved supercharges in their effective theories. 
In the case of $\U(N)$ gauge theory with $N$ fundamental chiral multiplets, 
we can show by examining the equations of motion for the fermions 
that there is one fermionic moduli parameter $\zeta^i_+$
for each bosonic moduli parameter $\phi^i$. 
The parameters $(\phi^i,\zeta^i_+)$ form 
a supermultiplet under the unbroken supersymmetry 
and transform in the same way under the torus action. 
According to the localization formula, 
if there is a supermultiplet $(\phi^i,\zeta^i_+)$
for which the weight of torus action is $\lambda$, 
its contribution to the vortex partition function is
$\lambda^{-2} \times \lambda = \lambda^{-1}$. 
Here $\lambda^{-2}$ and $\lambda$ are the contributions 
from the bosonic and fermionic part, respectively. 
Therefore, in the $\mathcal N=(2,2)$ $\U(N)$ gauge theory 
with $N$ fundamental chiral multiplets, 
the vortex partition function takes the form given in (\ref{part_func}).
In the next two examples, we will see that there exist 
fermionic zero modes which are not paired 
with dynamical bosonic zero modes. 

\subsection{Adding anti-fundamental matters}

In this section, we consider vortices in 2d $\mathcal N=(2,2)$ 
$\U(N)$ gauge theory with $N$ fundamental and 
$\widetilde N$ anti-fundamental chiral multiplets. 
In this case, a fermionic version of the moduli matrix
plays an important role.
Fermionic zero modes in 4d $\mathcal N=1$ theory 
in the case of $N=1$ and $\widetilde N=0$ was studied in 
\cite{Davis:1997bs}.
The simplest case with anti-fundamental matter, 
{\it i.e.}, the case of $N=\widetilde N=1$, 
was studied for a bosonic theory \cite{Penin:1996si} 
and an $\mathcal N=1$ supersymmetric theory \cite{Achucarro:2001ii}.

The relevant part of bosonic Lagrangian is
\beq
\mathcal L_b \, = \, \tr \bigg[ \D_\mu H (\D^\mu H)^\dagger + (\D_\mu \widetilde H)^\dagger \D^\mu \widetilde H - \frac{1}{2g^2} F_{\mu \nu} F^{\mu \nu} - \frac{g^2}{4} \left( H H^\dagger - \widetilde H^\dagger \widetilde H - v^2 \mathbf 1_N \right)^2 \bigg].
\eeq
Here, we have introduced the anti-fundamental field $\widetilde H$ 
($\widetilde N$-by-$N$ matrix)
in addition to the fields in the Lagrangian (\ref{boson_action01}).
In this case, the BPS equations are given by
\beq
0 &=& \partial_{\bar z} H + i A_{\bar z} H, \label{eq:BPS1} \\
0 &=& \partial_{\bar z} \widetilde H - i \widetilde H A_{\bar z}, \label{eq:BPS2} \\
0 &=& F_{12} - \frac{g^2}{2} ( H H^\dagger - \widetilde H^\dagger \widetilde H - v^2 \mathbf 1_N). \label{eq:BPS3}
\eeq
The general solution of the first two equations are written 
in terms of the moduli matrix not only for the fundamental field
but also for the anti-fundamental field,
\beq
A_{\bar z} = -i S^{-1} \partial_{\bar z} S, \hs{10} H = v \, S^{-1} H_0(z), \hs{10} \widetilde H = v \, \widetilde H_0(z) S.
\eeq
Note that both of the moduli matrices, $H_0(z)$ and $\widetilde H_0(z)$,
are holomorphic.

From the fundamental and anti-fundamental fields,
the mesonic gauge invariant quantity $M$ can be constructed as 
\beq
M ~\equiv~ \widetilde H H.
\eeq
For the BPS configuration, this invariant is a holomorphic function
\beq
M(z) = v^2 \, \widetilde H_0(z) H_0(z).
\eeq
Therefore, $M$ must be a constant matrix 
(otherwise $\lim_{z \rightarrow \infty} M = \infty$). 
Since $H_0(z)$ must be a rank-$N$ matrix, 
this condition is satisfied only when $M=0$, namely
\beq
\widetilde H_0(z) ~=~ \widetilde H ~=~ 0. 
\label{mesonic_condition}
\eeq
Thus, only the moduli matrix $H_0(z)$ for the fundamental scalars
can be non-trivial, so that 
the structure of the bosonic part of the moduli space 
is the same as in the case without the anti-fundamental matters 
discussed in section~\ref{sec:bosonic}.

Next, let us consider the fermionic part of the theory.  
Here, we use the convention of 4d $\mathcal N =1$ theories
for notational simplicity. 
The fermionic part of the Lagrangian reads
\beq
\mathcal L_f \, =  \, i \, \tr \bigg[ - \frac{1}{g^2} \lambda \sigma^\mu \D_\mu \overline \lambda + \overline \psi \overline \sigma^\mu \D_\mu \psi + \overline {\tilde \psi} \overline \sigma^\mu \D_\mu \tilde \psi - \left\{ \lambda \psi H^\dagger - \widetilde H^\dagger \tilde \psi \lambda - (h.c.) \right\} \bigg].
\eeq
The equations of motion coming from this part are given by
\beq
0 &=& \overline \sigma^\mu \D_\mu \psi + \overline \lambda H, \nonumber \\
0 &=& \overline \sigma^\mu \D_\mu \tilde \psi + \widetilde H \overline \lambda, \nonumber \\
0 &=& \sigma^\mu \D_\mu \overline \lambda + g^2 ( \psi H^\dagger + \widetilde H^\dagger \tilde \psi).
\eeq
These equations can be explicitly written 
in terms of the Weyl spinors in four dimensions.
Let $\psi_+, \psi_-,\cdots$ be the components of the Weyl spinors 
\beq
\psi_\alpha = \ba{c} \psi_+ \\ \psi_- \ea, \hs{10}
\tilde \psi_\alpha = \ba{c} \tilde \psi_+ \\ \tilde \psi_- \ea, \hs{10}
\overline \lambda^{\dot \alpha} = \ba{c} \phantom{-} \lambda_-^\dagger \\ - \lambda_+^\dagger \ea.
\eeq
In two dimensions $(\mu = 1,2)$, the covariant derivatives reduce to
\beq
\sigma^\mu \D_\mu = 
\ba{cc} 
0 & - 2 \D_z \\ 
- 2 \D_{\bar z} & 0 
\ea, \hs{10}
\overline \sigma^\mu \D_\mu = 
\ba{cc} 
0 & 2 \D_z \\ 
2 \D_{\bar z} & 0 
\ea.
\eeq
We now solve these equations by introducing fermionic holomorphic functions.
In the vortex background, the equations of motion for $\tilde \psi$ become\footnote{
The equations of motion for $\lambda$ and $\psi$ give 
the fermionic moduli $\zeta^i_+$ 
with which the bosonic moduli $\phi^i$ contained in $H_0(z)$ 
form the supermultiplets (see next subsection for details).
}
\beq
0 &=& \D_{\bar z} \tilde \psi_+ ~=~ \partial_{\bar z} \tilde \psi_+ - i \tilde \psi_+ A_{\bar z}, \\
0 &=& \D_z \tilde \psi_- ~=~ \partial_z \tilde \psi_- - i \tilde \psi_- A_z.
\eeq
Thus we have the following general solutions to these equations,
\beq
\tilde \psi_+ = \tilde \psi_{0+}(z) S, \hs{10} 
\tilde \psi_- = \tilde \psi_{0-} (\bar z) S^{\dagger -1}.
\eeq
Now let us consider the boundary condition for the fermionic zero modes. 
We simply assume that the fermionic zero modes vanish at the spatial infinity,
\beq
\lim_{|z| \rightarrow \infty} \tilde \psi_{\pm} = 0. 
\eeq 
Since the asymptotic form of $S$ for the vortex solution at the fixed point $(k_1,\cdots,k_N)$ is (see, e.g., \cite{Eto:2006pg})
\beq
S ~=~ \Big[ 1 + \mathcal O( e^{-g v |z|} ) \Big] \diag ( |z|^{k_1} , \cdots, |z|^{k_N} ) , 
\eeq
$\tilde \psi_+$ cannot have any regular solution satisfying 
the boundary condition. Therefore we have
\beq
\tilde \psi_{0+}(z) = 0.
\eeq 
On the other hand, the general form of the 
solution for $\tilde \psi_-$ yields
\beq
\tilde \psi_{0-} (\bar z) = 
\ba{ccc} 
\displaystyle \sum_{i=1}^{k_1} \zeta_{1 1, i} \, \bar z^{i-1} & \cdots & \displaystyle \sum_{i=1}^{k_N} \zeta_{1 N, i} \, \bar z^{i-1} \\
\vdots & \ddots & \vdots \\
\displaystyle \sum_{i=1}^{k_1} \zeta_{\widetilde N 1, i} \, \bar z^{i-1} & \cdots & \displaystyle \sum_{i=1}^{k_N} \zeta_{\widetilde N N, i} \, \bar z^{i-1} \ea,
\label{fermionic_mod_mat}
\eeq
where $\zeta_{lm,i}$ are Grassmann parameters. 
Therefore, the fermionic directions in the vortex moduli space are given by
\beq
\{ \tilde \psi_{0-} (\bar z) \} ~\cong~ \mathbb C^{\widetilde N} \otimes ( P_{k_1}[\bar z] \oplus \cdots \oplus P_{k_N}[\bar z] ).
\eeq
The torus action on $\tilde \psi_{0-}(\bar z)$ is defined by
\beq
\tilde \psi_{0-}(\bar z) ~\rightarrow~ T_\epsilon^{-1} T_m \tilde \psi_{0-}(T_{\epsilon}^{-1} \bar z) V_{\vec k}^\dagger.
\label{fermionic_torus_action}
\eeq
Here $T_m = \diag(e^{im_1},\cdots,e^{im_{\widetilde N}})$ 
stands for the torus action corresponding 
to the twisted masses for the anti-fundamental matters. 
The overall factor $T_\epsilon^{-1}$ is merely a convention 
and can be absorbed into the twisted masses. 
Thus the characters of the torus action on $\mathbb C^{\widetilde N}$
and $P_{k_1}[\bar z] \oplus \cdots \oplus P_{k_N}[\bar z]$ are simply
obtained from (\ref{fermionic_torus_action}),
\beq
\chi(\mathbb C^{\widetilde N}) ~=~ T_\epsilon^{-1} \sum_{f=1}^{\widetilde N} T_{m_f}, \hs{10}
\chi(P_{k_1}[\bar z] \oplus \cdots \oplus P_{k_N}[\bar z]) ~=~ \sum_{l=1}^N \sum_{i=1}^{k_l-1} T_\epsilon^i T_{a_l}.
\eeq
Therefore, we have the character for the fermionic tangent space 
\beq
\chi \left(\mathbb C^{\widetilde N} \otimes ( P_{k_1}[\bar z] \oplus
\cdots \oplus P_{k_N}[\bar z] ) \right) = \sum_{f=1}^{\widetilde N}
\sum_{l=1}^N \sum_{i=1}^{k_l-1} T_{m_f} T_\epsilon^{i-1} T_{a_l}. 
\label{antifund_char}
\eeq
This character correctly reproduces the previous result
\cite{Shadchin:2006yz,Bonelli:2011fq,Bonelli:2011wx}.
Note that similarly to the case of the bosonic zero modes 
discussed in section \ref{sec:bosonic}, 
we can easily extract this character (\ref{antifund_char}) 
from the coefficients $\zeta_{lm,i}$ 
in the fermionic moduli matrix (\ref{fermionic_mod_mat}), 
which are regarded as the fermionic coordinates of the vortex moduli space.
Since fermionic moduli parameters $\zeta_{lm,i}$ are not 
paired with any bosonic moduli, 
the contribution to the partition function 
from the anti-fundamental part is given by 
\begin{equation}
 Z_{\vec k}^{\rm antifund} 
  = \prod_{f=1}^{\tilde N} \prod_{l=1}^N \prod_{i=1}^{k_l-1}
  \left(
   m_f + a_l + (i-1) \epsilon
  \right) .
\end{equation}

\subsection{Adding an adjoint matter}
\begin{table}
\begin{center}
\begin{minipage}{0.45\textwidth}
\begin{center}
\begin{tabular}{c|ccc}
$ \U(1)_{\mathcal R} / \U(1)_J $ &  1 & 0 & $-1$ \\ \hline
0 & & $A_\mu$ & \\
1 & $\lambda_1$ & & $\lambda_2$ \\
2 & & $\varphi$ &  
\end{tabular}
\caption{vector multiplet}
\label{tab:vector}
\end{center}
\end{minipage}
\begin{minipage}{0.45\textwidth}
\begin{center}
\begin{tabular}{c|ccc}
$ \U(1)_{\mathcal R} / \U(1)_J $ &  1 & 0 & $-1$ \\ \hline
$-1$ & & $ \psi $ & \\
0 & $H$ & & $\widetilde H^\dagger$ \\
1 & & $\tilde \psi^\dagger$ &  
\end{tabular}
\caption{hypermultiplet}
\label{tab:hyper}
\end{center}
\end{minipage}
\end{center}
\end{table}
Next, let us consider 2d $\mathcal N = (2,2)$ $\U(N)$ gauge theory 
which can be obtained from 2d $\mathcal N =(4,4)$ $\U(N)$ gauge theory 
with $N$ fundamental hypermultiplets 
by adding a mass term for the adjoint chiral multiplet 
in the $\mathcal N =(4,4)$ vector multiplet. 
In this case, we have to deal with both bosonic and fermionic zero modes 
similarly to the case with the anti-fundamental chiral multiplets. 
See \cite{Shifman:2005st,Edalati:2007vk} for related analysis.

The $\mathcal N =(4,4)$ gauge theory consists of 
the vector and hypermutiplets, whose field contents 
(in 4-dimensional notion) and 
the $\U(1)_{\mathcal R}$ and $\U(1)_J \subset \SU(2)_R$ charges 
are summarized in Table \ref{tab:vector} and Table \ref{tab:hyper}. 
The BPS equations are the same as 
those in the previous section, 
that is, Eqs.\,(\ref{eq:BPS1})-(\ref{eq:BPS3}) with $\widetilde N = N$.
Therefore, the solution takes the form
\beq
A_{\bar z} = -i S^{-1} \partial_{\bar z} S, \hs{10} H = v \, S^{-1} H_0(z), \hs{10} \widetilde H = 0.
\eeq
The F-term constraint $\varphi H = 0$ implies that  
we have to choose $\varphi=0$ since $H$ has the maximal rank
in the BPS vortex configurations. 
As in the previous cases, 
all the bosonic moduli parameters are 
contained in the moduli matrix $H_0(z)$. 

On the other hand, the equations of motion for the fermions are
\beq
0 &=& \overline \sigma^\mu \D_\mu \psi + \overline \lambda_1 H + \overline \lambda_2 \widetilde H^\dagger + \varphi^\dagger \overline{\tilde \psi \,}, \phantom{\bigg[} \\
0 &=& \overline \sigma^\mu \D_\mu \tilde \psi + \widetilde H \overline \lambda_1 - H^\dagger \overline \lambda_2 - \overline{\psi} \varphi^\dagger, \phantom{\bigg[} \\
0 &=& \sigma^\mu \D_\mu \overline \lambda_1 + g^2 ( \psi H^\dagger + \widetilde H^\dagger \tilde \psi), \phantom{\bigg[}  \\
0 &=& \sigma^\mu \D_\mu \overline \lambda_2 + g^2 ( \psi \widetilde H - H \tilde \psi), \phantom{\bigg[} 
\eeq
In the BPS background, these equations become
\beq
\Delta
\ba{c} \psi_+ \\ \frac{i}{g} \lambda_{1+}^\dagger \ea = \Delta \ba{c} \tilde \psi_-^\dagger \\ \frac{i}{g} \lambda_{2-} \ea = 0,
\hs{10} 
\Delta^\dagger
\ba{c} \psi_- \\ \frac{i}{g} \lambda_{1-}^\dagger \ea = \Delta^\dagger \ba{c} \tilde \psi_+^\dagger \\ \frac{i}{g} \lambda_{2+} \ea = 0,
\label{eq:eq_f} 
\eeq
where $\Delta$ and $\Delta^\dagger$ are defined by
\beq
\Delta \equiv 
\ba{cc} 
i \D_{\bar z}^f & - \frac{g}{2} H_r \\
\frac{g}{2} H_r^\dagger & i \D_z^a  
\ea, \hs{10}
\Delta^\dagger \equiv 
\ba{cc} 
i \D_z^f & \frac{g}{2} H_r \\
- \frac{g}{2} H_r^\dagger & i \D_{\bar z}^a
\ea,
\eeq
where the subscript $r$ denotes the fact that $H$ 
acts as right multiplication and 
$\D_{\bar z}^f$ and $\D_z^a$ are covariant derivatives 
which act on the fundamental and adjoint fields, respectively.
Note that the linearized BPS equations for the bosonic zero modes 
are given by
\beq
\Delta \ba{c} \delta H \\ \delta A_{\bar z} \ea = 0 .
\eeq
The basis of the solutions of the linear differential equation 
$\Delta \Phi_i = 0$ are given by 
\beq
\Phi_i = \ba{cc} v \, S^{-1} \frac{\p}{\p \phi^i} H_0 \\ 0 \ea + \ba{cc}  i \omega_i H \\ - \frac{2}{g} \D_{\bar z} \omega_i \ea, \hs{10} \omega_i \equiv -i S^{-1} \left( \Omega \frac{\p}{\p \phi^i} \Omega^{-1} \right) S,
\eeq
where $\phi^i~(i=1,\cdots,{\rm dim}_{\C} \mathcal M_{N,k} =Nk)$
are bosonic moduli parameters contained in $H_0(z)$.
Therefore, the first two equations in (\ref{eq:eq_f}) give
the following two fermionic zero modes for each bosonic moduli parameter $\phi^i$:
\beq
\ba{c} \psi_+ \\ \frac{i}{g} \lambda_{1+}^\dagger \ea = \zeta^i_+ \Phi_i, \hs{10} 
\ba{c} \tilde \psi_-^\dagger \\ \frac{i}{g} \lambda_{2-} \ea = \zeta^i_- \Phi_i,
\eeq
where $\zeta^i_{\pm}$ are fermionic moduli parameters. 
On the other hand, it has been shown that 
there is no zero mode for $\Delta^\dagger$ \cite{Hanany:2003hp}. 
Now, let us consider the transformation property of 
the moduli parameters $(\phi^i,\,\zeta^i_+,\,\zeta^i_-)$
under the torus action. 
In this case , we have to include the $\U(1)_{\mathcal R-J}$ symmetry 
corresponding to the mass term for 
the adjoint scalar in the vector multiplet. 
The action of $\U(1)_{\mathcal R-J}$ on the original fields 
is given by
\beq
(H,\, \psi) \rightarrow e^{-\frac{i}{2} \mathfrak{m}} (H,\,\psi), \hs{10} (\widetilde H^\dagger,\,\tilde \psi^\dagger) \rightarrow e^{\frac{i}{2} \mathfrak{m}} (\widetilde H^\dagger,\,\tilde \psi^\dagger),
\label{eq:transf1}
\eeq
\beq
(A_\mu,\,\lambda_1) \rightarrow (A_\mu,\,\lambda_1), \hs{10} (\varphi,\lambda_2) \rightarrow e^{i\mathfrak{m}} (\varphi,\lambda_2).
\label{eq:transf2}
\eeq
Even in the presence of the adjoint mass $\mathfrak{m}$, 
the BPS solutions corresponding to the fixed points are not modified, 
that is, they are classified by $\vec k = (k_1,\cdots,k_N)$
and specified by the diagonal moduli matrices of the form (\ref{eq:fixed}). 
On the other hand, because of the transformation property of $H$, 
the matrix $V_{\vec k}$ given in (\ref{torus_and_V}) is modified as
\beq
V_{\vec k} \rightarrow \mathrm{diag} ( e^{-i(k_1\epsilon + a_1 + \mathfrak{m}/2)}, \cdots,
 e^{-i(k_N \epsilon + a_N + \mathfrak{m}/2)}).
\label{eq:transf3}
\eeq
In the case of $\mathfrak{m}=0$, 
the vortex configurations preserve four supercharges
since they are 1/2 BPS states in the $\mathcal N =(4,4)$ theory. 
The moduli parameters $(\phi^i,\,\zeta^i_+,\,\zeta^i_-)$ 
form supermultiplets (chiral multiplets) in the vortex effective theory 
and transform in the same way under the torus action. 
On the other hand, if the adjoint mass $\mathfrak{m}$ is non-zero, 
the multiplets $(\phi^i,\,\zeta^i_+,\,\zeta^i_-)$ are decomposed into
$(\phi^i,\,\zeta^i_+)$ and $\zeta^i_-$. 
We can show from (\ref{eq:transf1}), (\ref{eq:transf2}) 
and (\ref{eq:transf3}) that 
the weights of the multiplets $(\phi^i,\,\zeta^i_+)$ are not modified 
while the contributions from $\zeta^i_-$ are shifted by $\mathfrak{m}$. 
Therefore, the contribution to the vortex partition function 
from the fixed point $\vec k$ is given by
\beq
Z_{\vec k}^{\rm adj} = \prod_{l,m}^N \prod_{i=1}^{k_m} \frac{a_{ml}+(-k_l+i-1) \epsilon + \mathfrak{m}}{a_{ml}+(-k_l+i-1) \epsilon}.
\eeq
If we take the mass decoupling limit, $\mathfrak{m} \to \infty$, 
this reduces to the simple partition function given in (\ref{part_func}).
On the other hand, when we go back to $\N=(4,4)$ theory by taking the limit of
$\mathfrak{m} \to 0$, this contribution becomes trivial.
This situation is quite analogous to the four dimensional $\N = 2^*$ theory.

\section{Orbifold vortex partition function}
\label{sec:orbifold}

We then consider the vortex partition function for the orbifold theory
$\C/\Z_n$ \cite{Zhao:2011ke}.
Since supersymmetry on the two dimensional orbifold $\C/\Z_n$ is not
preserved \cite{Adams:2001sv}, availability of the localization formula
is questionable.
Thus the partition function we discuss in this section is a quite formal one.
Anyway it can be obtained in a similar manner to the orbifold instanton
partition function \cite{Fucito:2004ry}: it is given by considering the
invariant sector under the identification of the spatial coordinate as $z \sim
\omega z$, where $\omega = \exp (2\pi i/n)$ is the primitive $n$-th root
of unity.

Under the orbifold action $\Gamma = \Z_n$ the torus action  
behaves as
\begin{equation}
 T_\epsilon \longrightarrow \omega T_\epsilon, \qquad
 T_{a_l} \longrightarrow \omega^{p_l} T_{a_l}, \qquad
 T_{m_f} \longrightarrow \omega^{p_f} T_{m_f},
\end{equation}
where $p_l$ and $p_f$ are parameters, satisfying $0 \le p_l, p_f \le n-1$.
They characterize the irreducible representation of $\Gamma=\Z_n$ in the
flavor space.
This is regarded as the twisted boundary condition, a kind of holonomy,
for the flavor space.
Thus contribution to the character for the fundamental
chiral multiplets (\ref{character}) is modified as
\begin{equation}
 T_{a_{ml}} T_\epsilon^{-k_l+i-1}
  \longrightarrow
  \omega^{p_{ml}-k_l+i-1} T_{a_{ml}} T_\epsilon^{-k_l+i-1}.
\end{equation} 
If the extra $\omega$-factor vanishes, it is invariant under the
orbifold action $\Gamma=\Z_n$, and thus contributes to the character.
This means the $\Gamma$-invariant sector is given by
\begin{equation}
 p_{ml}-k_l+i-1 \equiv 0 \quad (\mbox{mod}~n).
  \label{inv_sec}
\end{equation}
The orbifold vortex partition function for $N$ fundamental chiral
multiplets is given by a product over the $\Gamma$-invariant sector,
\begin{equation}
  Z_{\vec{k},\Gamma} = \prod_{l,m}^N \prod_{\mbox{\scriptsize $\Gamma$-inv.}} 
  \frac{1}{a_{ml} + (-k_l + i - 1) \epsilon}.
  \label{part_func_orb}
\end{equation}

The orbifold partition function for the anti-fundamental matter is given
in a similar manner.
The corresponding contribution to the character (\ref{antifund_char}) is
modified as
\begin{equation}
 T_{m_f} T_\epsilon^{i-1} T_{a_l}
  \longrightarrow
 \omega^{p_f+p_l+i-1} T_{m_f} T_\epsilon^{i-1} T_{a_l} .
\end{equation}
The orbifold invariant sector for this part yields
\begin{equation}
 p_f + p_l + i - 1 \equiv 0 \quad (\mbox{mod}~n) .
  \label{inv_sec_antifund}
\end{equation}
Therefore the orbifold partition function for the anti-fundamental
matter is given by
\begin{equation}
 Z_{\vec k, \Gamma}^{\rm antifund} 
  = \prod_{f=1}^{\tilde N} \prod_{l=1}^N 
  \prod_{\mbox{\scriptsize{$\Gamma$}-inv.}}
  \left(
   m_f + a_l + (i-1) \epsilon
  \right) .
\end{equation}
Here the product is taken over the $\Gamma$-invariant sector defined in
(\ref{inv_sec_antifund}).
Similarly the orbifold partition function for the adjoint matter theory is also obtained with the same $\Gamma$-invariant sector
(\ref{inv_sec}),
\begin{equation}
 Z_{\vec k, \Gamma}^{\rm adj} = \prod_{l,m}^N 
  \prod_{\mbox{\scriptsize{$\Gamma$-inv.}}} 
  \frac{a_{ml}+(-k_l+i-1) \epsilon + \mathfrak{m}}{a_{ml}+(-k_l+i-1) \epsilon}.
\end{equation}
This partition function is reduced to 
(\ref{part_func_orb}) in the decoupling limit, $\mathfrak{m} \to \infty$, 
and becomes trivial in the massless limit, $\mathfrak{m} \to 0$.

This orbifold partition function is also derived from the moduli matrix method.
The moduli matrix approach to the orbifold theory has been investigated
in \cite{Kimura:2011wh}, especially in the absence of the twisted mass
terms and so on.
When we consider the twisted mass terms, the moduli matrix has to
satisfy the following condition,
\begin{equation}
  H_0(\omega z) = \Omega H_0(z) \tilde \Omega^\dag, 
   \label{orb_cond}
\end{equation}
\begin{equation}
  \Omega = \mathrm{diag}(\omega^{k_1+p_1}, \cdots, \omega^{k_N+p_N}),
   \qquad
   \tilde \Omega = \mathrm{diag} (\omega^{p_1}, \cdots, \omega^{p_N}).
\end{equation}
From both sides of (\ref{orb_cond}) the neighborhood around the fixed
point yields
\begin{eqnarray}
  \left(\delta H_0(\omega z)\right)_{lm} & = & \sum_{j=1}^{k_m} \omega^{j-1}
   c_{lm,j} z^{j-1},
   \nonumber \\
 \left(\Omega \delta H_0(z) \tilde \Omega \right)_{lm} & = & \sum_{j=1}^{k_m}
  \omega^{k_l+p_l-p_m} c_{lm,j} z^{j-1},
\end{eqnarray}
This means the coefficient has to satisfy $c_{lm,j}=0$ unless
$\omega^{j-1}=\omega^{k_l-p_{ml}}$, which is equivalent to the condition
(\ref{inv_sec}).
% In this case, to satisfy the boundary condition for the moduli matrix,
% we have to remove factors whose powers are different modulo $n$ from
% that of the diagonal component in the same row.
%This condition is explicitly written as
%\begin{equation}
% c_{lm,j} \left\{\begin{array}{cccc}
%	   \not = 0 & \mbox{for} & j-1-k_l \equiv 0 & (\mbox{mod}~n) \\
%	   = 0 & \mbox{for} & j-1-k_l \not\equiv 0 & (\mbox{mod}~n) \\
%		 \end{array}\right. .
%\end{equation}
%This is just corresponding to the $\Gamma$-invariant sector (\ref{inv_sec}).
Thus we can obtain the orbifold partition function 
in the similar way as the usual case of vortices on $\C$.
We can apply the same argument to the cases 
with the anti-fundamental and adjoint matters.

\section{Summary and discussion}
\label{sec:summary}

In this paper, we have investigated the vortex partition function 
from the field theoretical point of view 
by using the moduli matrix approach.
Since the moduli matrix itself is interpreted as the moduli space coordinates, 
one can easily see how the isometry acts on the tangent space 
at the fixed points in the vortex moduli space.
The corresponding character has been also written 
in terms of the moduli matrix, 
and thus we have consistently derived the vortex
partition function in a field theoretical way.

We have mainly dealt with 
2d $\N=(2,2)$ $\U(N)$ theories 
with the twisted masses for the chiral multiplets, 
which break the $\SU(N)$ color-flavor diagonal group 
into the maximal torus $\U(1)^{N-1}$.
There is a certain variety of the matter contents: 
we have considered the cases with $\widetilde{N}$ anti-fundamental, 
and the adjoint matters in addition to 
$N_{\rm F}=N$ fundamental chiral multiplets.
Due to the partial breaking of the supersymmetry, 
not only the standard bosonic, 
but also the fermionic moduli matrix also contributes 
to the vortex partition function.

We have then considered the vortex partition function 
for the orbifold $\C/\Z_n$.
By studying the consistency conditions 
for the bosonic and fermionic moduli matrices in the orbifold theory, 
we have similarly derived the character of the tangent space 
at the fixed points, 
and thus the orbifold vortex partition function 
from the field theoretical perspective.
These conditions are regarded as natural extensions 
of the original one, which is discussed in \cite{Kimura:2011wh}.

We now comment on some possibilities of future works.
\if0
It seems interesting to investigate the fermionic moduli matrix, 
which is used to obtain the additional contribution 
from the anti-fundamental and adjoint matters to the partition function.
In theories with eight supercharges 
all fermionic zero modes are dictated by their bosonic partners, 
while they can be independent of bosonic zero modes 
and the fermionic moduli matrix becomes non-trivial 
in the case of four supercharges.
It is worth discussing the vortex moduli space structure 
from the viewpoint of the fermionic moduli matrix, 
and its effect on the dynamics of vortices and so on.
\fi
In this paper, we have studied vortices with additional matters 
in the anti-fundamental and adjoint representations, 
in addition to $N$ fundamental matters. 
When there are more {\it fundamental} matter fields, $N_{\rm F}>N$, 
vortices are called {\it semi-local} \cite{Vachaspati:1991dz,Achucarro:1999it}.
Non-Abelian semi-local vortices were studied 
in \cite{Shifman:2006kd,Eto:2007yv}. 
Since the vortex counting in the semi-local case was studied 
for the $\U(1)$ gauge theory \cite{Dimofte:2010tz}, 
an extension to non-Abelian semi-local vortices should be explored. 
In those cases, we can also discuss the relation between 
the vortex partition functions and 
the non-perturbative twisted superpotentials 
which determine the BPS mass spectra in $\mathcal N =(2,2)$ theories 
\cite{Dorey:1998yh,Dorey:1999zk}. 
It explains the concidence of the BPS mass spectra 
in 2d and 4d gauge theories \cite{Shifman:2004dr,Hanany:2004ea},
since semi-local vortices (sigma model instantons) 
in the non-Abelian vortex world-sheet 
can be identified with Yang-Mills instantons 
in the bulk point of view \cite{Hanany:2004ea,Eto:2004rz,Fujimori:2008ee}.
Along this line, a new 4d/2d correspondence 
has been recently proposed \cite{Dorey:2011pa,Chen:2011sj}.

A natural extension of the result obtained in this paper 
is application to other gauge group theories.
The moduli matrix approach to the theory 
with the gauge symmetry of $G \times \U(1)$ such as $G = \SO, \mathrm{USp}$ 
is investigated in \cite{Eto:2009bg,Eto:2008qw,Eto:2011cv}.
The merit of field theory approach is 
that the construction of vortices is available 
even for arbitrary groups $G$ \cite{Eto:2008yi} 
for which D-brane (K\"ahler quotient) construction are not known.
We can apply the method developed in this paper to such theories,
and obtain vortex partition functions.
The instanton partition functions are also given for 
the theories with various gauge group \cite{Nekrasov:2004vw,Keller:2011ek}, 
which can be written in terms of the corresponding root systems.
It is expected that the vortex partition functions can also be written 
in a similar manner.
It seems also important to study partition functions 
for quiver gauge theories \cite{Bonelli:2011wx}, 
and their relation to the AGT correspondence \cite{Alday:2009aq}.

We can consider other two dimensional spaces, 
e.g. cylinder \cite{Eto:2006mz}, 
torus \cite{Eto:2007aw,Lozano:2007bk}, 
hyperbolic surfaces \cite{Manton:2010wu} and 
more general Riemann surfaces
\cite{Popov:2008gw,Baptista:2008ex,Baptista:2010rv,Manton:2010mj}.
In particular, we can apply the localization formula 
to the partition function on $S^2$ 
in a similar way to the case of $S^4$ \cite{Pestun:2007rz}.
Furthermore, it will be interesting to 
consider various defects which partially preserve the supersymmetry 
and investigate its relation to the domain-wall partition function \cite{Ohta:2007ji}.

\subsection*{Acknowledgments} 
T.~K. is supported by Grant-in-Aid for JSPS Fellows.
The work of M.~N. is supported in part by 
Grant-in Aid for Scientific Research (No.~23740226) 
and by the ``Topological Quantum Phenomena'' 
Grant-in Aid for Scientific Research 
on Innovative Areas (No.~23103515)  
from the Ministry of Education, Culture, Sports, Science and Technology 
(MEXT) of Japan.

%%%%%%%%%  Rerefence  %%%%%%%%%

%\bibliographystyle{JHEP}
%$\bibliographystyle{apsrev4-1}
\bibliographystyle{ytphys}

\bibliography{vortex}

\end{document}